\documentclass[graybox]{svmult}

\usepackage{mathptmx}       % selects Times Roman as basic font
\usepackage{helvet}         % selects Helvetica as sans-serif font
\usepackage{courier}        % selects Courier as typewriter font
\usepackage{type1cm}        % activate if the above 3 fonts are
\usepackage{makeidx}         % allows index generation
\usepackage{graphicx}        % standard LaTeX graphics tool
\usepackage{multicol}        % used for the two-column index
\usepackage[bottom]{footmisc}% places footnotes at page bottom

\usepackage{cite}

\usepackage[normalem]{ulem}
\usepackage[none]{hyphenat}
\RequirePackage{color}
\definecolor{MyDarkGreen}{rgb}{0.02,0.60,0.06}

\makeindex             % used for the subject index
\begin{document}

\title*{Analyses of a Virtual World }
\author{Yurij Holovatch, Olesya Mryglod, Michael Szell, and Stefan Thurner}
\institute{Yurij Holovatch \at Institute for Condensed Matter
Physics, National Academy of Sciences of Ukraine, 79011 Lviv, Ukraine,
\email{hol@icmp.lviv.ua} \and Olesya Mryglod \at Institute for
Condensed Matter Physics, National Academy of Sciences of Ukraine, 79011
Lviv, Ukraine, \email{olesya@icmp.lviv.ua} \and Michael Szell \at
Center for Complex Network Research, Northeastern University, 02115
Boston, Massachusetts, USA, \email{m.szell@neu.edu} \and Stefan
Thurner \at Section for Science of Complex Systems, Medical
University of Vienna, Vienna, Austria; Santa Fe Institute, Santa Fe,
NM 87501, USA; IIASA, Schlossplatz 1, A-2361 Laxenburg, Austria,
\email{stefan.thurner@meduniwien.ac.at} }
\maketitle

\abstract*{ We present an overview of a series of results
obtained from the analysis of human behavior in a virtual
environment. We focus on the massive multiplayer online game (MMOG)
{\em Pardus} which has a worldwide participant base of more than 400,000
registered players.
We provide  evidence for striking
statistical similarities between social structures and human-action
dynamics in the real and  virtual worlds.
In this sense MMOGs
provide an extraordinary way for accurate and falsifiable studies of
social phenomena.
We further discuss possibilities to apply methods
and concepts developed in the course of these studies to analyse oral and
written narratives.}

\abstract{ We present an overview of a series of results
obtained from the analysis of human behavior in a virtual
environment. We focus on the massive multiplayer online game (MMOG)
{\em Pardus} which has a worldwide participant base of more than 400,000
registered players.
We provide  evidence for striking
statistical similarities between social structures and human-action
dynamics in  real and  virtual worlds.
In this sense MMOGs
provide an extraordinary way for accurate and falsifiable studies of
social phenomena.
We further discuss possibilities to apply methods
and concepts developed  in the course of these studies to analyse oral and
written narratives.}

\section{Introduction}
\label{sec:1}

Quantitative approaches in social sciences and humanities have
benefited greatly from the introduction of advanced information
technologies. These allow one to accumulate and store a huge amount
of data, as well as to enable its effective processing.
Computer-based communication technologies have led to the formation
of virtual societies, and these societies have themselves become the
subjects of research. In this Chapter, we demonstrate some results
obtained through analyses of human behavior in a  Massive
multiplayer online game (MMOG) (Castronova 2005). Playing such games
has become one of the largest collective human activities in the
world; at present hundreds of millions of people participate in such
activities including, for example, approximately 10 million who are
registered for the most popular MMOG \emph{World of Warcraft}
(Statista 2015). In turn, the records of activity of players in
MMOGs provide extraordinary opportunities for quantitative analyses
of social phenomena with levels of accuracy that approach those of
the natural sciences. The results we discuss below were obtained
from a series of analyses of the MMOG {\em Pardus}. Since it was
launched in 2004, the {\em Pardus} game served as a unique testing
ground to measure different observables that characterize
inhabitants of the virtual world and in this way to obtain clues
also on complex social processes taking place in the
real\footnote{We use the word ``real'' due to lack of a better term.
Certainly human behavior, emotions, and decisions in online worlds
are as ``real'' as in the offline world -- they might only be biased
differently depending on the context.} world (Szell and Thurner
2010, Szell et al. 2010, Szell et al. 2012, Thurner et al. 2012,
Shell and Thurner 2013, Klimek and Thurner 2013, Corominas-Murtra et
al. 2014, Fuchs and Thurner 2014, Fuchs et al. 2014, Sinatra and
Szell 2014).

The reasons for the appearance of a chapter on a multiplayer online
world in a book devoted to complexity\--science approaches to oral
and written narratives may not be obvious at first sight.
Comparative mythology, folktales and epic literature which are the
main issues in this book have little to do with the virtual world of
\emph{Pardus}. However, a more careful comparison reveals a number
of common features and potentially transferrable analytical tools.
In both cases, one treats narrative or virtual characters in a
manner similar to how sociologists treat real social groups, with an
aim to quantify properties of such groups. See e.g. Stiller et al.
(2003), Mac Carron and Kenna (2012), Mac Carron and Kenna (2013) and
references therein. In such studies, quantitative analyses put
comparison and classification of different narratives on a solid
basis. A similar goal is pursued by the analysis of actions of
virtual characters (players) of an MMOG.

Although the societies of an MMOG  and of a narrative are to some
extent mirrors of the real world, they reflect it in different ways.
In an MMOG, each individual  is the character controlled by a player
(i.e., an avatar or graphical representation of the user). In a
narrative, the individual is a character in a story. The narrative
is created with the intention to be perceived by a reader and it
carries a personal contribution of a writer. Life in an MMOG evolves
as a complex system and is driven by  numerous interactions between
players. Here we discuss some results of analyses of the virtual
world and methods used to obtain them which, we hope, might be also
useful in future analyses of the world of narratives. Moreover,
analysis of life in a synthetic world serves as a tool to learn more
about human behavior in the real world.

We discuss  the application of complex-network concepts to uncover the diversity of social interactions in a society.
We pay particular attention to  how multidimensional graphs, wherein nodes may be connected by more than one type of edge or link, contribute to the formation of
different interconnected  {\emph{multiplex}} networks.
We demonstrate how one can test traditional social-dynamical hypotheses which apply
to virtual societies too, bringing to the fore intrinsic similarities between virtual and real worlds.
In addition, we analyse the evolution of social networks in time through a  first
 analysis of dynamical features of multi-level human activity (sequences of human actions of different types).
This study of multi-level human activity in Section 3 can be seen as a dynamic
counterpart to static multiplex network analysis presented in the following section.

\section{Database and Networks}
\label{II}

\emph{Pardus} is a browser-based MMOG  played since September 2004.
It is an open-ended game with no explicit winning purpose and a
worldwide player base of more than 400,000 registered participants
(Pardus 2015, Szell and Thurner 2010). The game has a science
fiction setting and each player controls one character. The
characters act within a virtual world, making up their own goals and
interacting with the  social environment which is self-organized.
The game features three different universes: \emph{Orion},
\emph{Artemis}, and \emph{Pegasus}. Each universe has a fixed start
date but no scheduled end date. The results we discuss here concern
the \emph{Artemis universe}, selected for study because it has most
active players and because its data set is most complete. Artemis
was opened on June 10, 2007 and at the time of this study was
inhabited by several thousand active characters.

Each character in the game is a pilot who owns a spacecraft, travels
in the universe and is able to perform a number of activities of
different types, such as communication, trade, attack, establishing or breaking friendships
or enmities, etc.
 Since we focus on  social features, we make use of records about the following activities of each character in the game:
\begin{itemize}
\item sending private messages  from one player to another (communication, C);
\item attacking other players or their belongings (attack, A);
\item trading or giving gifts (trade, T);
\item indicating friends by adding their names to a friend list (F);
\item indicating enemies by adding their names to an enemy list  (E);
\item removing friends from the friend list (D);
\item removing enemies from the enemy list (X);
\item placing a bounty on other players (B).
\end{itemize}
The overall number of actions performed by the characters during
1,238 consecutive days of observation was $N=8,373,209$ [for a
detailed description of the database see Szell and Thurner (2010),
Mryglod et al. (2015)].

A straightforward way of mapping the \emph{Pardus} society onto a
complex network is to associate  nodes with individual characters. A
link between two nodes represents an action that took place between
the corresponding pair of characters. Every action type (from the
above list) is directed; it is initiated by one character and
directed towards another. Given the different possible actions, one
arrives at a set of directed networks where links in each network
correspond to actions of certain types.
We define  the in- and
out- degrees as the number of incoming and outgoing links  that a
given node has, and we denote these by $k^{\rm in}$ and $k^{\rm
out}$ respectively. When we don't specify whether we are dealing
with in- or out-degrees, or if it doesn't matter (when the network
is not directed), we use the generic symbol $k$ for the degree
instead. Data are available with a one-second resolution,
making possible a refined analysis of dynamical features of the
virtual society. With the data on activities of each player to
hand, one can construct networks of social interactions
at each instant in time and follow their evolution.

\begin{figure}[tb]
    \begin{center}
\includegraphics[width=0.7\textwidth]{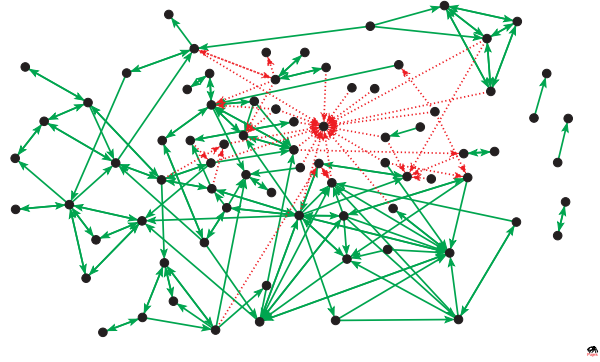}
    \end{center}
    \caption{Friendship (green, solid) and enemy (red, dashed) relations on day 445 between 78 randomly selected game characters.
        The diagram is taken from Szell and Thurner (2010). An animated evolution of this network can be seen at: {\ttfamily{http://www.youtube.com/user/complexsystemsvienna}}
    }
    \label{fig:1}
\end{figure}

\begin{figure*}[tb]
    \begin{center}
        \includegraphics[width=0.9\textwidth]{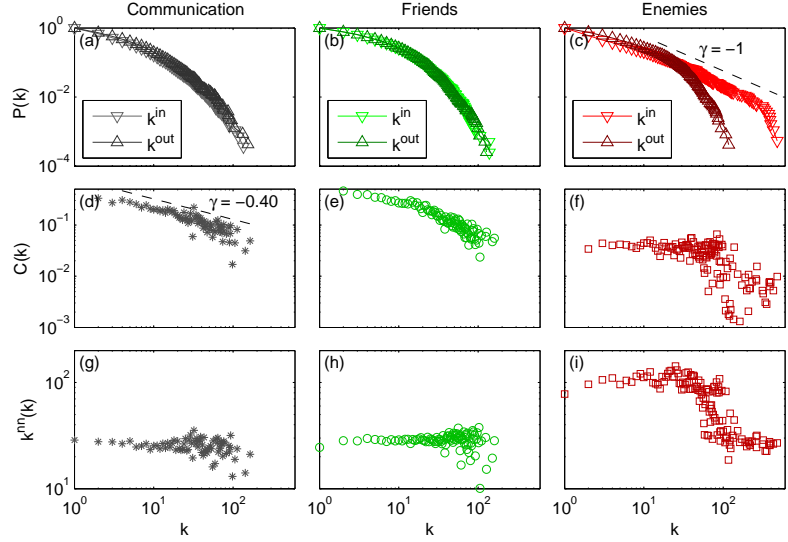}
    \end{center}
\caption{Cumulative degree distributions of (a) communication (C),
(b) friendship (F)  and (c) enemy (E) networks; clustering
coefficient $C$ as a function of degree $k$ for the (d) C, (e) F and
(f) E networks; nearest neighbor degree $k^{\mathrm{nn}}$ versus
degree of the (g) C, (h) F, and (i) E networks. Fits to power laws
($\sim k^{\gamma}$) are indicated by dashed lines, when appropriate.
All distributions are shown as for 2008-09-01, the picture is taken
from Szell and Thurner (2010).} \label{fig:2}
\end{figure*}

As an example, in Fig. \ref{fig:1} we show networks of friendship
and enemy relations on the 445th day (01 September 2008) between 78
randomly selected characters. One can measure the basic network
properties, track their evolution over time  (Szell and Thurner
2010) and quantify correlations between properties of networks of
different types  (Szell et al. 2010). In Fig. \ref{fig:2} we show
several features of the communication (C), friendship (F), and enemy
(E) networks, measured during the same day in the virtual world. We
display three measures: the cumulative degree distribution $P(k)$,
the clustering coefficient $C(k)$ and the mean degree $k^{\rm
nn}(k)$. The first of these, $P(k)$, is the probability that the
degree  $k_i$ of a randomly selected node $i$ is at least as large
as a given value $k$ ($k_i\geq k$). The second measure, $C(k)$, is
defined in the following manner. One first forms the ratio of the
number of links which {\emph{actually}} exist between $i$'s
neighbours and the number of all {\emph{possible}} links between
them. If node $i$ has $k_i$ neighbours, each of these  can be linked
to $k_i-1$ other neighbours of node $i$. The total number of
potential links between $i$'s neighbours is therefore
$k_i(k_i-1)/2$, having divided by two to deal with the overcounting
induced by each link being shared by two nodes. If $y_i$ is the
actual number of links between the neighbours of node $i$, then we
define the clustering of the $i$th node to be
\begin{equation}\label{1}
C_i=\frac{2y_i}{k_i(k_i-1)}\, .
\end{equation}
%Thus the clustering associated with node $i$ is the number of links which {\emph{actually}} exist between $i$'s neighbours as a fraction of the number of links which {\emph{could}} exist between them.
Taking the average of the $C_i$'s over all nodes for which $k_i=k$
gives the mean clustering $C(k)$. This is the average clustering
associated with nodes of degree $k$. The third and final  measure in
Fig. \ref{fig:2}, $k^{\rm nn}(k)$, is the mean degree of the nearest
neighbours of nodes which themselves have degree $k$. These and
other basic network properties of the \emph{Pardus} society are
discussed in detail in Szell and Thurner (2010) and further analyzed
in numerous publications (Szell et al. 2010, Szell et al. 2012,
Thurner et al. 2012, Shell and Thurner 2013, Klimek and Thurner
2013, Corominas-Murtra et al. 2014, Fuchs and Thurner 2014, Fuchs et
al. 2014, Sinatra and Szell 2014, Mryglod et al. 2015).

Fig.2 tells us that there are a number of characteristics that distinguish networks of different types.
Each plot is on a double-logarithmic scale so that any power-laws present would show up as straight-line segments.
For example, comparing the properties of the communication, friendship, and enemy
networks, one finds that only in the latter case can the cumulative
node degree distribution be approximated by a power law.
This is evidenced in Fig.\ref{fig:2}{\bf c} where the fit indicates that
$P(k) \sim k^{\gamma}$ with $\gamma = -1$.
The corresponding plots for the communication and friendship cases, Fig.\ref{fig:2}{\bf a} and Fig.\ref{fig:2}{\bf b}, respectively, are evidently not described by power-laws.

Fig.~\ref{fig:2}{\bf d} and Fig.~\ref{fig:2}{\bf e} show that the clustering coefficients $C(k)$
for the communicaton and friendship networks exhibit  clear downward trends as $k$
increases, whereas Fig.~\ref{fig:2}{\bf f} shows that the clustering
coefficients are to a large extent independent of $k$ for the enemy networks, at least for large values of $k$.
That the  same can be said about the degrees $k^{\rm{nn}}$ is evident from Figs.~\ref{fig:2}{\bf g} to {\bf i}.

These observations can be complemented by  examining the behavior of
the \emph{linking probability} $p(k)$ as function of the degree $k$
(not shown in Fig.\ref{fig:2}). This is the probability of a new
node  connecting to an existing node with degree $k$. By fitting it
with a power-law function of the type $p(k)\sim k^{\alpha}$, we can
try to understand how the network grows as new nodes are added. If
$\alpha$ is positive, it means that new nodes prefer to attach to
nodes which have higher degrees. In the case where $\alpha=1$, which
signals a linear dependency, this phenomenon is known as
{\emph{preferential attachment.}} The following values for the
exponents have been reported for in-degrees: $\alpha\simeq 0.62$ for
the friendship  and $\alpha\simeq 0.90$ for the enemy network. We
refer to Szell et al. (2010a) for further discussions on this topic.

These quantitative observations are important because they allow one
to conclude that there are intrinsic differences in the network
formation processes for the C, F and E networks. In particular, in a
typical preferential-attachment mechanism, links   attach to nodes
according to how many links these nodes already have; high-degree
nodes receive more new neighbours than their low-degree
counterparts. The resulting node degree distribution decays as a
power law, the linking probability increases linearly with $k$ (i.e.
$\alpha=1$) and the clustering coefficient $C(k)$ is uniform as a
function of $k$  (Barab\'asi and Albert 1999). None of these three
properties  holds for the communication or friendship networks, but
all  are roughly satisfied for the enemy network. This suggests that
the preferential attachment scenario is mainly relevant for the
latter. In other words, the more enemies a character has, the more
likely they are to accrue more enemies but the same is not true for
friends or for communication.

The above discussion around Fig.~\ref{fig:2} focuses on the
differences between the various network types. There are
similarities too. The following common features have been observed
for \emph{Pardus} networks of different types (Szell and Thurner
2010): (i) their average node degrees grow over time; (ii) the mean
shortest path length decreases with time; (iii)  networks such as
communication, trade and friendship, are reciprocal (individuals
tend to reciprocate connections); (iv) but networks such as enmity,
attack and bounty are not. Properties (i) and (ii) signal that the
network becomes more dense as time evolves while properties (iii)
and (iv) show that networks with positive connotations tend to be
reciprocal while those with negative connotations are not.

\begin{figure*}[tb]
    \begin{center}
        \includegraphics[width=0.9\textwidth]{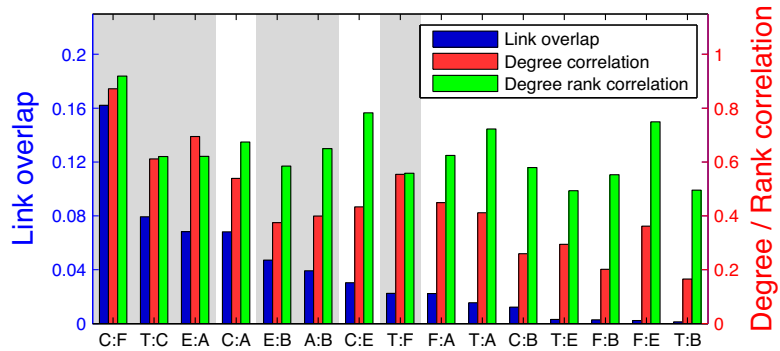}
    \end{center}
\caption{Link overlap, degree correlation, and degree rank
correlation for different pairs of networks.   Pairs of equal
connotation (positive-positive or negative-negative) are marked with
a gray background.  See Szell et al. (2010) for more explanations,
where from the picture has been taken. } \label{fig:3}
\end{figure*}

The example network shown in Fig. \ref{fig:1} is a part of a
multiplex network (Wasserman and Faust 1994), that consists of a set
of characters  that are joined by different types of links,
corresponding to different types of social relations (recorded as
different actions, in the case of the \emph{Pardus} game). It is
well established by now that multiplexity plays an essential role in
network organization. Indeed, an interplay between different social
relations, expressed as an interaction of links of different types,
may lead to new levels of complexity. To quantify the relations
between multiplex network layers, a thorough analysis for the
\emph{Pardus} society has focused on the \emph{link overlap} and
correlations between node degrees between the different network
layers (Szell et al. 2010).

The \emph{link overlap} quantifies the interaction between two networks by
measuring the tendency that edges are simultaneously are present in both
networks.
It is defined by the so-called \emph{Jaccard coefficient}, which is a similarity score between two sets of elements.
It is defined as the size of the intersection of the sets divided by the size
of their union.
It is therefore a global measure which ranges in size from zero (no overlap) to 1 (100\% overlap).

The correlation between node degrees (or their ranks), on the other hand, measures the extent to which degrees of agents in one type of network correlate with degrees of the same agents in the other one.
If the correlation between node degrees in two different networks is close to 1, players who have many  links in one network have many links in the other one and vice versa.
In Fig. \ref{fig:3} we show these quantities for different pairs of networks.
One sees that pairs of the same connotation (positive-positive or
negative-negative)   typically have high overlaps, whereas this is not
the case for pairs of  opposite connotation.
Moreover, low values
of the degree correlations indicate that hubs in one network are not
necessarily hubs in another. This demonstrates the tendency of
individuals to play different roles in different networks.

Properties of the multiplex network vary for different types of
players. A recent analysis of gender-specific differences has shown
(Shell and Thurner 2013) that females and males manage their social
networks in substantially different ways. In particular, on the
individual level, females perform better economically and are less
inclined to take risks than males. Males reciprocate friendship
requests from females faster than vice versa and hesitate to
reciprocate hostile actions of females. On the network level,
females have more communication partners, who are themselves less
connected than partners of males. Cooperative links between males
are under-represented, reflecting competition for resources among
males.

Analysis of the \emph{Pardus} universe also allows one to quantify
to what extent classical sociological hypotheses hold up in a
virtual world. Such analyses recently enabled us to propose two
approximate social laws in communication networks (Szell and Thurner
2010). These findings were made in the course of testing
Granovetter's \emph{Weak Ties Hy\-po\-the\-sis} (Granovetter 1973),
which suggests casual acquaintanceships link communities in an
essential way. This means that weak links are important to hold the
network together - it is the weak links, not the strong ones, that
tend to form the ties between distinct sets of nodes. To explain
this quantitatively, we require three new concepts. Firstly, the
{\emph{overlap}} $O$ is the fraction of common neighbours between
two neighbouring nodes; if  $A$ and $B$ represent the sets of
neighbours of two nodes, the overlap is the number of nodes in  that
$A$ and $B$ have in common ($A \cap B$) divided by the total number
in $A$ and $B$ taken together ($A \cup B$). This is a local measure,
distinct from the global link overlap discussed earlier.
 Secondly, the {\emph{link-betweenness centrality}}, $b$, is
the ratio of the number of shortest paths between two nodes that
contain a given link to the total number of shortest paths between
these nodes. Thirdly we need the {\emph{weight}} $w$ of a link
joining  two nodes. This is also a local quantity. For the
communication network, the weight of a link between two nodes
corresponds to the number of private messages sent between two
individuals these nodes represent.

The stronger the connection between two individuals, the more
similar is their local environment, and vice versa. Therefore we
expect that the overlap is an increasing function of weight.
Analysis of the structure of communication networks in the
\emph{Pardus} world has indeed revealed that, on average,  the
overlap $O$ related to  the weight $w$ of a link joining  two given
nodes increases as
\begin{equation}\label{soc1}
O(w)\sim\sqrt[3]{w} \, .
\end{equation}
To understand how this connects with Granovetter's hypothesis, consider two sets of nodes: one set involving Node A and one including Node B.
If the set of nodes connected to Node A is very distinct from the set of nodes connected to Node B, then the overlap $O$ between the two sets is low (it is zero if they are completely distinct).
If $O$ is small, Eq.(\ref{soc1}) tells us that $w$ is also a small number.
This means that the weight $w$ between nodes A and B is small on average.
This low weight corresponds to the notion of casual relationship.
Thus Eq.(2) quantifies Granovetter's hypothesis - it tells us that light-weight relationships are essential to bind distinct sets of nodes.

Another way to quantify the Weak Ties Hypothesis is to check the
behavior of the overlap $O$ as a function of link-betweenness
centrality, $b$.
If the hypothesis is valid, shortest connections between two sets of nodes are forced to go through the weak links that connect them.
In other words, low overlap corresponds to high betweenness.
The obtained dependency was indeed found to
be a decreasing function of the explicit form
\begin{equation}\label{soc2}
O(b)\sim \frac{1}{\sqrt b} \, ,
\end{equation}
which supports the hypothesis.

Other social hypotheses that were tested and confirmed for the
\emph{Pardus} networks concern triadic closure and network
densification (Granovetter 1973). The triadic closure conjecture
follows balance considerations (Heider 1946) and reflects the
property among three nodes A, B, and C in a social network,  that if
node pairs A-B and A-C are linked by strong ties, there tends to be
a weak or strong tie between the node pair B-C. The phenomenon of
triadic closure (Rapoport, 1953) states that individuals are driven
to reduce the cognitive dissonance caused by the absence of a link
in the (unclosed) triad. Because of this the triad in which there
exist strong ties between all three subjects A, B and C should
appear in a higher than expected frequency. Network densification
(i.e. shrinking of its diameter and growing of average degrees with
a span of time) is an aging effect that has been observed recently
in many growing networks (Leskovec et al. 2007). Observation of
similar effects for {\em Pardus} networks serves as one more
argument about universal features of this phenomenon. It is  worth
noting one feature known in the real world that is also reflected in
the virtual society. It concerns the number of people with whom one
can maintain stable social relationships, given humans' limited
cognitive capacities. This is the so-called ``Dunbar'' number
(Dunbar, 1993). See also Kenna and Berche (2010) for a mathematical
basis for the upper limit of group sizes.
 A prominent feature of the plots in
Fig. \ref{fig:2} is that the maximal out-degree of networks
represented there is limited by $k^{\rm out}\simeq 150$, a value
conjectured to be the maximal number of stable relationships humans
can comfortably maintain (Dunbar 1993).\footnote{Editors'note: See Robin Dunbar's chapter in this volume.}

Results discussed so far give a quantitative
description of the \emph{Pardus} society based on a network perspective. We
 argued how networks of different social
interactions arise and evolve, how they interact with one another,
what are the observables that describe their properties and what are
their implications for life in a virtual world. Although
dynamical features of network evolution were also analyzed here, we did
not address so far the question of temporal structure of human
actions. We now ask if there exist regularities that govern temporal
behavior of characters in a virtual world and if so, do they
resemble those in the real world? Some answers to this question will
be given in the following section.

\section{Human Multi-Level Activity}
\label{III}

The lives of humans can be viewed as sequences of different actions.
Some of these are performed on a regular basis; others have strong
stochastic components. Some actions are performed frequently; others
are carried out sporadically. One associated quantity of interest is
the time-lapses between such events - the ``inter-event'' times.
Many early models which were used to study the inter-event time
distribution of human action sequences were based on the assumption
that such actions are performed randomly in time. In  the simplest
cases, these are described statistically by a Poisson process. This
assumption suggests that times between actions of the same
individual are independent and distributed exponentially. Models of
such kind are still being used, however there appears to be an
accumulation of evidence  that distribution functions characterizing
sequences of different human actions in time are highly non-trivial
[see Barab\'asi (2005), Oliveira and Barab\'asi (2005), Vazquez et
al. (2006), Malmgren et al. (2009), Goh and Barab\'{a}si (2008), Wu
et al. (2010), Jo et al. (2012), Yasseri et al. (2012a), Yasseri et
al. (2012b) and references therein]. An inhomogeneous bursty
distribution of human actions influences their temporal statistics
and often is associated with power laws. Such conclusions were made
on the basis of observing different types of single human action
such as writing letters, checking out books in libraries, writing
e-mails, web browsing, and many more. Analysis of temporal features
of the performance of actions of different types, which we call a
multi-level human activity,  still remains an open challenge. The
main problem here is the obvious difficulty in accessing reliable
and statistically relevant databases of records of various forms of
human activity.

The clear advantage of our data set on the multi-level activity of
characters in the \emph{Pardus} world is that it is based on the
analysis of behavior of thousands of characters across several
years, and that it concerns various types of actions. In this sense
it can be considered as the dynamic counterpart of static multiplex
network analysis. The main outcome of this study is given in this
section. The interested reader is referred to Mryglod et al. (2015)
for a more extensive report.

\begin{figure*}[tb]
    \begin{center}
        \includegraphics[width=0.9\textwidth]{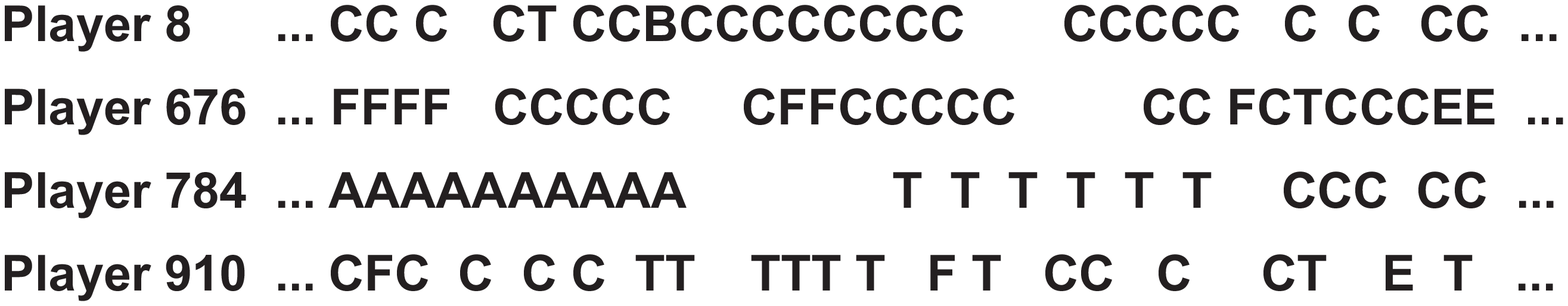}
    \end{center}
    \caption{Short segment of action sequences, performed by four players.
Different actions are shown by different letters as  explained in
section \ref{II}. \label{fig:4}}
    \end{figure*}

\begin{figure*}[tb]
    \begin{center}
        \includegraphics[width=0.48\textwidth]{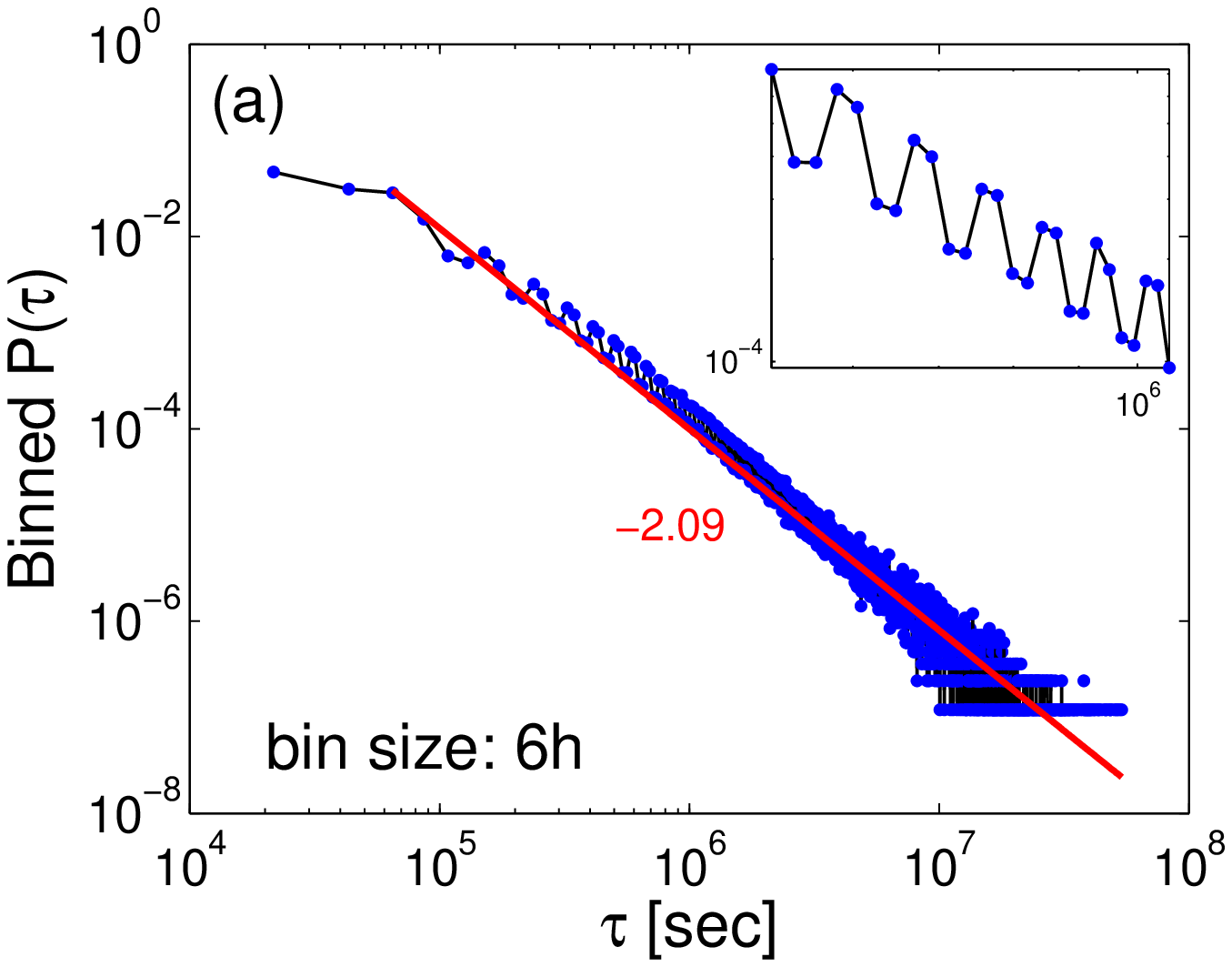}
        \includegraphics[width=0.48\textwidth]{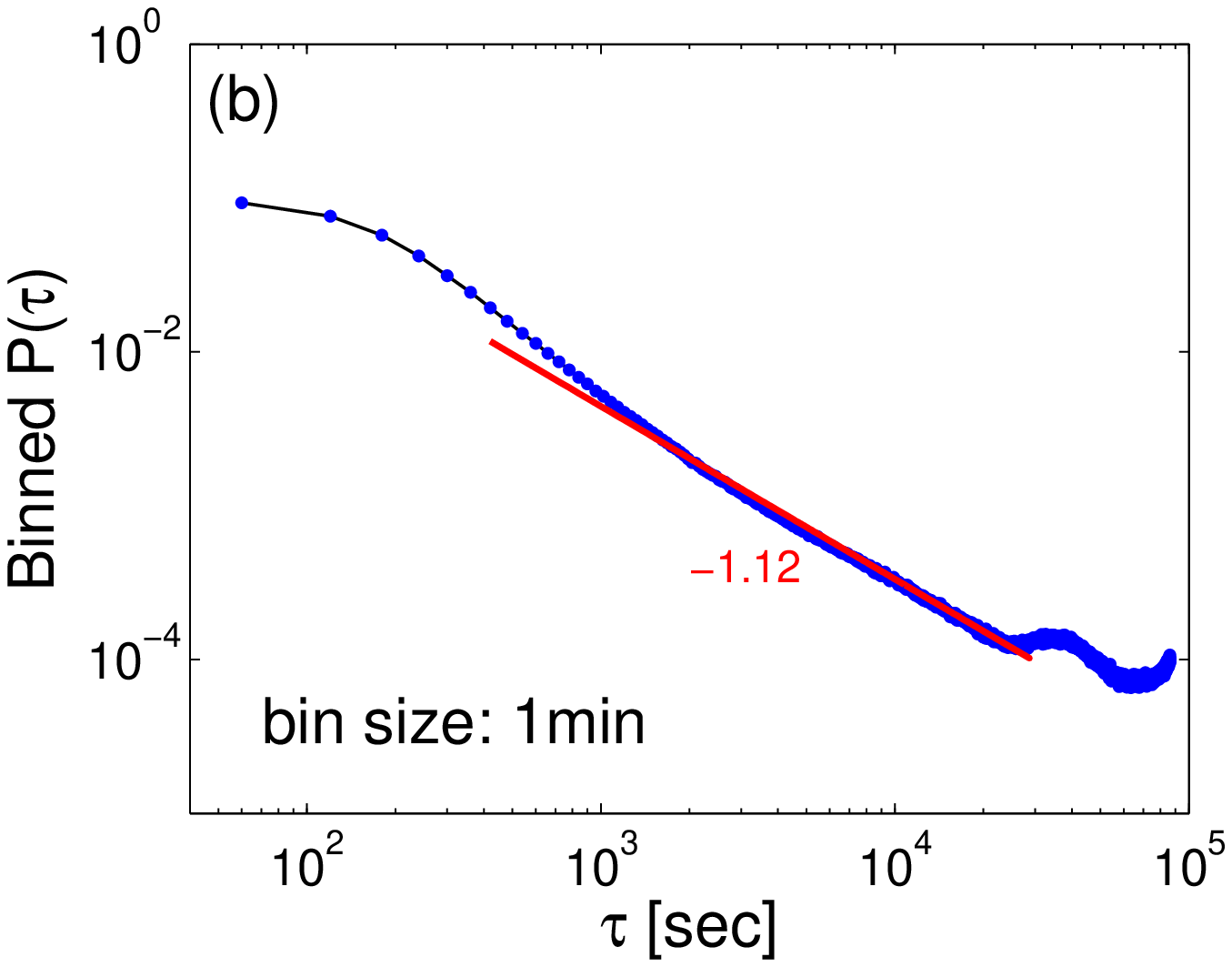}
    \end{center}
    \caption{Distribution of the inter-event times $\tau$ for all players who performed at least 50 actions.
(a) entire observation period (1,238 days), bin size is 6
hours=$21,600$ sec. (b) first 24 hours, bin size is 1 min. Inset:
same as (a) for six days. Circadian rhythms are clearly visible. The
picture is taken from Mryglod et al. (2015). \label{fig:5}}
    \end{figure*}

Fig. \ref{fig:4} shows four segments of action sequences, performed
by four \emph{Pardus} players. Different actions are shown by different
letters as explained at the beginning of section \ref{II}. The
 times for each action have been recorded which allows us to
analyze peculiarities of temporal behavior of each player during the
whole observation time (for the results shown below it is equal to
1238 days since  10 June 2007 when the \emph{Artemis universe} was
opened). Further, we can assemble a general picture of temporal
behavior of all players. Below we concentrate on the statistics of
inter-event times $\tau$, i.e. the time intervals between two
consecutive actions of the same player. In Fig. \ref{fig:5} we show
the distribution functions of the  inter-event time $\tau$ for all
actions\footnote{Here we take into account all actions as listed at
the beginning of section \ref{II}, discarding for technical reasons
the 'bounties' B.} of all players who performed at least 50 actions,
considering players with fewer actions being not representative. As
it can be seen from Figs. \ref{fig:5}{\bf a}, {\bf b},  the
distributions follow approximate power laws, the numerical value of
the exponent depending on the chosen bin size. A prominent feature
of the plots is that they manifest a fine structure: one observes
regular patterns of various periodicity when the plots are
considered on a smaller scale. The emergence of periodic patterns is
known to be an inherent feature of human activity (Jo et al. 2012,
Yasseri et al. 2012a, Malmgren et al. 2009). In our case it can be
naturally explained by circadian and active working day cycles as
well as by peculiarities of performing different actions (Mryglod et
al. 2015).

The power-law behavior of inter-event time distribution functions
signals the bursty nature of human dynamics (Barab\'asi 2005,
Oliveira and Barab\'asi 2005). One of the variables that is used to
quantify burstiness is the so-called burstiness index (Goh and
Barab\'{a}si 2008), $B$, defined as (Jo et al. 2012, Yasseri et al.
2012b)
\begin{equation}\label{burst}
B\equiv \frac{\sigma-m}{\sigma+m}\, ,
\end{equation}
where $m$ and $\sigma$ represent the average inter-event time  and
the standard deviation, respectively.
\begin{figure*}[b]
\centering
\sidecaption
\includegraphics[width=6.0cm]{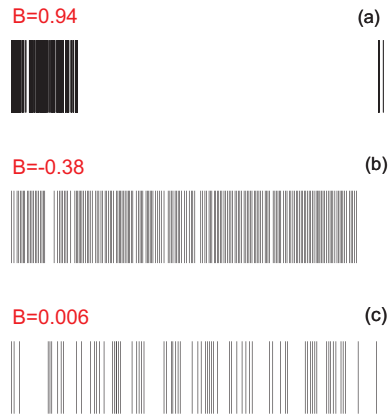}
\caption{Action streams of three players, each  with different values of
burstiness $B$. Lines mark times of  executed actions, the distance
between lines is the inter-event time. \label{fig:6}}
\end{figure*}
As follows from definition in Eq.~(\ref{burst}), a value of
$B\approx -1$ characterizes regular patterns. Random Poisson process
with a fixed event rate yields $B\approx 0$. In Fig. \ref{fig:6} we
show several action streams of individual players with different
values of burstiness. Panels ({\bf a}), ({\bf b}), and ({\bf c})
display action streams where $B$ is maximal, minimal, and close to
zero. We find that the average value of burstiness for all actions
of all players is $\overline{B}\simeq 0.53$. This feature of the
virtual world can be compared with the  burstiness values
characterizing real world activities of mobile communication
$\overline{B}\simeq0.2$ (Jo et al. 2012), and Wikipedia
editing\footnote{In this case, the events correspond to consequent
edits of Wikipedia articles.} $\overline{B}\sim0.6$ (Yasseri et al.
2012b). Note that, similar to the real world, the burstiness is
action-specific in the virtual world too. We find that burstiness of
attacks has larger values than for communication. This illustrates
the intuitive understanding of the nature of these actions: attacks
(A) appear highly clustered within short time intervals, while
communication (C) is more uniformly distributed over time.

Another inherent feature of the multi-level human activity in the
\emph{Pardus} world is that time distributions found there are
action-specific: each type of actions has a particular and
characteristic distribution. This fact is far from trivial, the
claim being that the  decay of the inter-event time distributions
might serve as a distinguishing feature of action type. Similar to
physics, where one can classify life times of unstable elements by
decay constants that uniquely characterize each element, one can
quantify the decay of action-specific inter-event time distributions
by the decay constants that uniquely characterize different types of
actions too. Leaving the detailed discussion of this fact to a
separate publication (Mryglod et al. 2015) where the inverse
cumulative distributions $P(\geq \tau)$ of inter-event times were
analyzed and the numerical values of the decay constants can be
found, we mention here, that the overall behavior of players is
characterized by three different time scales: (i) an immediate
reaction ($\tau$ does not exceed several minutes), (ii) an early day
($\tau$ is less then 8 hours), (iii) a late day ($\tau$ is between 8
and 24 hours). At long times (more then several months) an
exponential cut off becomes apparent, whereas for very short times
(time scale (i)) all distributions have a similar tendency to decay
very fast: the short inter-event times are typical for most of the
actions. The scales (ii) and (iii) bring about specific features of
different actions and allow to classify them. We found that on scale
(ii) the inverse cumulative distributions of inter-event times of
every specific action are best approximated by a power law:
\begin{equation}\label{power}
P(\geq \tau) \sim \tau^{-\alpha}\, ,
\end{equation}
while on scale (iii) the decay is of exponential form:
\begin{equation}\label{exp}
P(\geq \tau) \sim \exp\, (-\tau/\tau_0)\, .
\end{equation}
Constants $\alpha$ and $\tau_0$  allow to discriminate between
actions of different type and to quantify them in the unique way.

\begin{figure*}[tb]
    \begin{center}
        \includegraphics[width=0.8\textwidth]{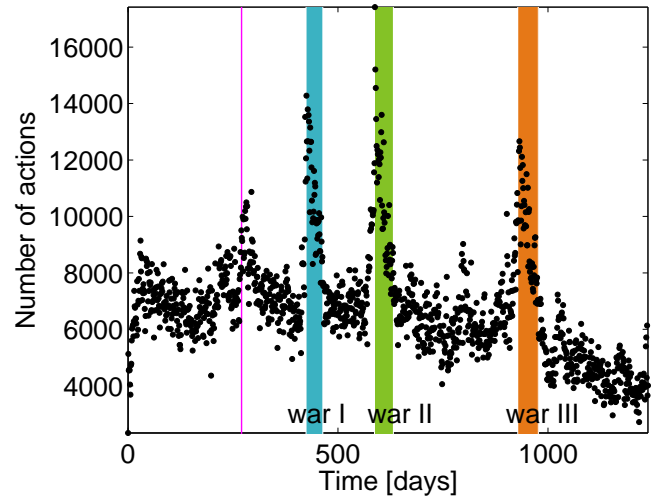}
            \end{center}
    \caption{Numbers of actions per day of all types over time.
There are four pronounced peaks in the player activity that
corresponds to specific  events that happened in the virtual world during
the observation period: the three coloured vertical stripes indicate
war periods, the thin vertical line indicates the introduction of a major
new game feature. \label{fig:7}}
    \end{figure*}

Before finishing this chapter we comment on the global dynamics and
activity patterns that can be observed in the \emph{Pardus} world.
In Fig. \ref{fig:7} we show how actions are distributed over time.
One can see four pronounced peaks in the players' activities. They
correspond to specific  events that happened in the virtual world
during the observation period: the three coloured vertical stripes
in the figure indicate war periods, the vertical line indicates the
introduction of a major new game feature. Besides an obvious
conclusion about the increase of activity during war periods,
changes in action-specific dynamics are observed (e.g.,
intensification of attacks and communication). Another question of
interest was to check whether the changes in player activity might
serve as precursors of coming wars (or precursors of the  end of
war). Interaction and coexistence of different social relations are
important to describe conflicts in social systems (Bohorquez et al.
2009, Lim et al. 2007, Clauset and Gleditsch 2012). Assuming that a
war in a virtual world emerges and finishes as a result of a complex
process of social interaction, it is tempting to ask about details
of early and late stages of this process. Our attempts to use
cross-correlation analysis for finding potential lead effects of
player activity patterns on the onset of war have not (yet) lead to
conclusive answers (Mryglod et al. 2015).

\section{Conclusions and Outlook}
\label{IV}

To what extent is the behavior of characters in a virtual world
similar to human behavior in the real world? To answer this
question, one has to compare quantitative behavioral features in
both worlds. Results obtained in the analysis of player behavior in
the MMOG \emph{Pardus} provide solid evidence for the existence of
certain similarities of social structure and human dynamics in the
real and the virtual worlds. These similarities concern certain
types of social networks, their growth patterns, the validity of
major sociological hypotheses, gender specific dynamics, etc. In
this sense, MMOGs provide an extraordinary opportunity for an
accurate analyses of social phenomena and falsifiable hypotheses.

Let us return  to the question about the comparison of social
activities in a virtual world and in the (written or oral)
narrative. Two properties of the virtual society are obvious: (i)
its structure changes over time, (ii) every element of the society
(every character) acts in time. To what extent might the study of
these properties be useful for similar analyses of narratives? In
section \ref{II} we have shown how property (i) is covered within
the network formalism.  In addition, one may use inter-event time
distributions to quantify property (ii). To this end, the
application of a multi-level human activity formalism may be useful,
as outlined in section \ref{III}. In the analysis of social networks
of narratives, usually the resulting static networks of all acting
characters are studied (Mac Carron and Kenna 2012). In principle,
one can get access to their evolution too, introducing the time-line
via counts of narrative subunits [pages or chapters (Mac Carron and
Kenna 2013), or appearance of new actors (Dunbar 1993)]. In this
sense, property (i) is accessed in the narrative analysis too. We
are not aware of analyses of property (ii) for a narrative. Although
such a problem statement might be interesting, its realisation,
besides obvious difficulties in introducing a coherent and
self-consistent time line, will meet difficulties of separating
dynamics caused by the evolution of a subject and a style of
presentation.

\begin{acknowledgement}
We want to express our thanks to the Editors of the book, Ralph
Kenna, M\'air\'in Mac Carron, and P\'adraig Mac Carron, for the
invitation to write this chapter and for useful suggestions and to
Anita Wanjek for helpful comments on the manuscript. This work was
supported in part by the 7th FP, IRSES project No.~612707 Dynamics
of and in Complex Systems (DIONICOS) and by the COST Action TD1210
Analyzing the dynamics of information and knowledge landscapes
(KNOWSCAPE). ST acknowledges support by the EU FP7 project LASAGNE
no. 318132.
\end{acknowledgement}

\end{document}